\RequirePackage{fix-cm}
\documentclass[twocolumn,epjc3]{svjour3}  
\smartqed  
\RequirePackage{graphicx}
\RequirePackage{bm}
\RequirePackage{mathrsfs}
\RequirePackage{amsmath}
\RequirePackage{cancel}
\RequirePackage{amssymb}
\RequirePackage{url}
\RequirePackage{cases}
\RequirePackage{tcolorbox}
\journalname{Eur. Phys. J. C}
\begin{document}
\sloppy
\title{Attractive Heaviside-Maxwellian (Vector) Gravity from Special Relativity and Quantum Field Theory}

\author{Harihar Behera\thanksref{e1,addr1}
        \and N. Barik\thanksref{e2,addr2}} 
\thankstext{e1}{e-mail: behera.hh@gmail.com}
\thankstext{e2}{e-mail: dr.nbarik@gmail.com}
\institute{Physics Department, BIET Higher Secondary School, Govindpur, Dhenkanal-759001, Odisha, India \label{addr1}
           \and
           Department of Physics, Utkal University, Vani Vihar, Bhubaneswar-751004, Odisha, India \label{addr2}}
\maketitle
\begin{abstract} 
Adopting two independent approaches (a) Lorentz-invariance of physical laws  and (b)  local phase invariance of quantum field theory applied to the Dirac Lagrangian for massive electrically neutral Dirac particles, we rediscovered the fundamental field equations of Heaviside Gravity (HG) of 1893 and Maxwellian Gravity (MG), which look different from each other due to a sign difference in some terms of their respective field equations. However, they are shown to represent two mathematical representations of a single physical theory of vector gravity that we name here as Heaviside-Maxwellian Gravity (HMG),  in which the speed of gravitational waves in vacuum is uniquely found to be equal to the speed of light in vacuum. We also corrected a sign error in Heaviside's speculative gravitational analogue of the Lorentz force law. This spin-1 HMG is shown to produce attractive force between like masses under static condition, contrary to the prevalent view of field theorists. Galileo's law of universality of free fall is a consequence of HMG, without any initial assumption of the equality of gravitational mass with velocity-dependent mass.  We also note a new set of Lorentz-Maxwell's equations having the same physical effects as the standard set - a byproduct of our present study.

\end{abstract}
\section{Introduction}
\label{intro}
Many field theorists, like Gupta \cite{1}, Feynman \cite{2}\footnote{On page 30 of ref. \cite{2}, Feynman noted: ``A spin-1 theory would be essentially the same as electrodynamics. There is nothing to forbid the existence of two spin-1 fields, but gravity can't be one of them, because one consequence of the spin 1 is that likes repel, and un-likes attract. This is in fact a property of all odd-spin theories; conversely, it is also found that even spins lead to attractive forces, so that we need to consider only spins 0 and 2, and perhaps 4 if 2 fails; there is no need to work out the more complicated theories until the simpler ones are found inadequate."}, Zee \cite{3} and Gasperini \cite{4}\footnote{On page 27, Gasperini noted: 
``A correct description of gravity in the relativistic regime thus requires an appropriate generalization of Newton’s theory. Which kind of generalization? A natural answer seems to be suggested by the close formal analogy existing between the Newton force among static masses and the Coulomb electrostatic force among electric charges. In the same way as the Coulomb potential corresponds to the fourth component of the electromagnetic vector potential, the Newton potential might correspond to the component of a four-vector, and the relativistic gravitational interaction might be represented by an appropriate vector field, in close analogy with the electromagnetic theory.\\
Such an attractive speculation, however, has to be immediately discarded for a very simple reason: vector-like interactions produce repulsive static interactions between sources of the same sign, while - as is well known - the static gravitational interaction between masses of the same sign is attractive.” }, to name a few, have rejected spin-1 vector theory of gravity on the ground that if gravitation is described by a vector field theory like Maxwell's electromagnetic theory, then vector-like interactions will produce repulsive static interactions between sources of the same sign, while - according to Newton's gravitational theory - the static gravitational interaction between masses of the same sign is attractive. Misner,Thorne and Wheeler (MTW)\cite{5}, in their ``Exercises on flat space-rime theories of gravity", suggested an action functional for a possible vector theory of gravity within the framework of special relativity and asked the reader to find it to be deficient in that there is no bending of light, incorrect value for the perihelion advance of Mercury and gravitational waves carry negative energy in a vector theory. Nevertheless, there have been several studies on vector gravitational field theory (reviewed here in Section 2) ever since Maxwell's \cite{6} first unsucessful attempt in 1865 and later Heaviside's \cite{7,8,9,10,11,12,13} successful theoretical formulation of the fundamental field equations of a vector gravitational theory, called Heaviside Gravity (HG), which we derive here following two independent approaches: (a) using the Lorentz invariance of physical laws and (b) using the principle of local gauge invariance of quantum field theory as applied to a massive electrically neutral Dirac spin-$1/2$ Fermion. However, Heaviside's speculative gravitational analogue of the Lorentz force law had a sign error, whose correction we report for the first time in this paper through our derivation. 
Alongside, using the above two approaches we also derived the fundamental equations of Maxwellian Gravity (MG) \cite{14} which we show to be physically equivalent to HG despite the appearance of some sign differences in certain terms of their respective equations. Because of our establishment of the equivalence between HG and MG, we named the resulting vector theory here as Heaviside-Maxwellian Gravity (HMG). Since the explanations of the classical tests of general relativity (GR) pointed out by MTW within the framework of vector gravity now exist in the literature \cite{15,16,17}, the main aim of this paper is to show the attractive interaction between two static (positive) masses in a vector theory of gravity, contrary to the prevalent view of the field theorists. Moreover, we suggest a Lagrangian (density) for this vector field theory of gravity in which gravitational waves carry positive energy. \\
This paper is organized as follows. Section 2 details a review of vector gravitational theory. In Section 3, the fundamental equations of HMG are derived using the Lorentz invariance of physical laws by adopting Behera and Naik's approach to Maxwellian gravity (MG)\cite{14}, wherein Galileo's law of universality of free fall is a consequence of the theory, without any initial assumption of the equality of gravitational mass with velocity dependent inertial mass - whose violation is demonstrated in a relativistic thought experiment that resolves Eddington's ``gravitational mass ambiguity" \cite{18} (stated here in Sec.3). The new findings in Section 3, not explicitly shown by Behera and Naik \cite{14} are (i) the relativistic rediscovery of Heaviside Gravity (HG) of 1893 \cite{7,8,9,10,11,12,13}, (ii) the establishment of the physical equivalence of HG with MG\cite{14} and thereby naming the resulting theory here as Heaviside-Maxwellian Gravity (HMG), (iii) a correction to gravitational analogue of the Lorentz force law speculated by Heaviside (iv) Suggestion of a Lagrangian that reproduces all of HMG with gravitational waves carrying positive energy. In Section 4, we follow the usual procedure of quantum electrodynamics (in flat space-time) starting with the free Dirac Lagrangian, the requirement of local phase invariance now applied to massive electrically neutral Dirac particles having rest mass $m_0$ to find a Lagrangian that generates all of gravitodynamcis of HMG and specifies the current produced by massive Dirac particles. Spin-1 graviton is described in Section 5; while in Section 6, we show the attraction between two static (positive) masses in the frame-work of HMG. In Section 6, we note our conclusions. 
\section{Vector Gravity: A Brief Review}   
By recognizing the striking structural similairy of Newton's law of gravitational interaction between two masses and Coulomb's law of electrical (or magnetic) interaction between two charges (or magnetic poles) and also their fundamental differences, J. C.Maxwell \cite{6}, in sect. 82 of his great 1865 paper, \emph{A Dynamical Theory of the Electromagnetic Field}, made \emph{a note on the attraction of gravitation}, in which he considered whether Newtonian gravity could be extended to a form similar to the form of electromagnetic theory - a vector field theory - where the fields in a medium possess intrinsic energy. As a first step in this line of thought, Maxwell calculated the intrinsic energy $U_g$ of the static gravitational field at any place around gravitating bodies:
\begin{equation}\label{eq:1}
U_g = C - C^\prime \int_{\text{All space}} \mathbf{g}^2 d^3x
\end{equation}
where $C$ and $C^\prime$ are two positive constants and $\mathbf{g}$ is the gravitational field intensity at the place. If we assume that energy is essentially positive\footnote{Which is not true if one considers gravitostatic field energy only. In fact following the electrostatic field energy calculation (see for example, Griffiths's \emph{Introduction to Electrodynamics}) one obtains $U_g = - \frac{1}{8\pi G}\int_{\text{All space}} \mathbf{g}^2 d^3x$. Thus one can set $C = 0$ and $C^\prime = \frac{1}{8\pi G}$ in Eq. \eqref{eq:1}. The value of $U_g$ calculated by this field theoretical method by using \eqref{eq:1} with $C = 0$ and $C^\prime = \frac{1}{8\pi G}$, for a spherical body of mass $M$, radius $R$ with uniform mass density within the body's volume, turns out as $U_g = -\frac{3}{5}\frac{GM^2}{R}$, which is the correct Newtonian (non-field-theoretic) result.}, as Maxwell did\footnote{By stating, ``As energy is essentially positive it is impossible for any part of space to have negative intrinsic energy."}, then the constant C must have a value greater than $C^\prime \mathbf{g}^2$, where $g$ is the greatest value of the gravitational field at any place of the universe: and hence at any place where $|\mathbf{g}|= 0$, the intrinsic energy must have an enormously great value. Being dissatisfied with this result, Maxwell, concluded his note on gravitation by stating, \emph{``As I am unable to understand in what way a medium can possess such properties, I can not go any further in this direction in searching for the cause of gravitation''}. \\
\indent
The first written record of a vector gravitational theory was made by Oliver 
Heaviside \cite{7,8,9,10,11,12,13} in 1893. Studying by electromagnetic analogy, he found a
 set of four field equations for gravity akin to Maxwell's equations of electromagnetism
  representing what we call Heaviside Gravity (HG). The gravitational field equations of
   HG, as we recently notice in Heaviside's original work, appear in the following
    Maxwellian form (written here in our notation).\\\\
{\bf Field Equations of Heaviside Gravity (HG):}
\begin{subequations}\label{eq:2}
\begin{equation}
\label{eq:2:1}
\mathbf{\nabla}\cdot\mathbf{g}\,=\,-\,4\pi G \rho_0\,=\,-\,\rho_0/\epsilon_{0g},
\end{equation}
\begin{equation}
\label{eq:2:2}
\mathbf{\nabla}\times\mathbf{b} = \frac{4\pi G}{c_g^2} \mathbf{j} -\frac{1}{c_g^2}\frac{\partial \mathbf{g} }{\partial t} = \mu_{0g}\mathbf{j} - \frac{1}{c_g^2}\frac{\partial \mathbf{g} }{\partial t},
\end{equation} 
\begin{equation}
\label{eq:2:3}
\mathbf{\nabla}\cdot\mathbf{b}\,=\,0,
\end{equation} 
\begin{equation}
\label{eq:2:4}
\mathbf{\nabla}\times\mathbf{g}\,=\,\frac{\partial \mathbf{b}}{\partial t}.
\end{equation} 
where 
\begin{equation}
\label{eq:2:5}
\epsilon_{0g} = \frac{1}{4\pi G}, \quad \mu_{0g} = \frac{4\pi G}{c_g^2} \quad \Rightarrow \quad c_g = \frac{1}{\sqrt{\epsilon_{0g}\mu_{0g}}}
\end{equation}
\end{subequations}
with $c_g$ representing the the speed of gravitational waves in vacuum, which might well be the speed of light $c$ in vacuum as Heaviside thought it, $\rho_0$ is the ordinary (rest) mass density, $\mathbf{j} = \rho_0\mathbf{v}$ is the mass current density ($\mathbf{v}$ is velocity) and by electromagnetic analogy, $\mathbf{b}$ is called the gravitomagnetic field, the Newtonian gravitational field $\mathbf{g}$ is called the gravitoelectric  field, $\epsilon_{0g}$ is called the gravitelectric (or gravitic) permitivity of vaccum and $\mu_{0g}$ is called the gravitomagnetic permeability of vacuum. To complete the dynamic picture, in a subsequent paper (Part II) \cite{7,8,9,10,11,12,13} Heaviside speculated a gravitational analogue of Lorentz force law in the following form:
\begin{equation}
\label{eq:3}
\mathbf{F}_{gL}^{HG} = m_0\frac{d\mathbf{v}}{dt} = m_0\mathbf{g} + m_0\mathbf{v}\times\mathbf{b} \qquad \text{(speculated),}
\end{equation}
to calculate the effect of the $\mathbf{b}$ field (particularly due to the motion of the Sun through the cosmic aether) on Earth's orbit around the Sun. As will be shown in this paper, the correct gravito-Lorentz force law for HG should be of the form:
\begin{equation}
\label{eq:4}
\mathbf{F}_{gL}^{HG} = m_0\frac{d\mathbf{v}}{dt} = m_0\mathbf{g} - m_0\mathbf{v}\times\mathbf{b} \qquad \text{(corrected).}
\end{equation}
However, Heaviside by considering Eq. \eqref{eq:3} calculated the precession of Earth's orbit around the Sun and concluded that this effect was small enough to have gone unnoticed thus far, and therefore offered no contradiction to his hypothesis that gravitational effects propagate at the speed of light. Surprisingly, Heaviside seemed to be unaware of the long history of measurements of the precession of Mercury's orbit as noted by McDonald \cite{19}, who reported Heaviside's gravitational equations (in our present notation) as given below under the name Maxwellian Gravity. \\\\
{\bf Field Equations of Maxwellian Gravity (MG):}
\begin{subequations}\label{eq:5}
\begin{equation}
\label{eq:5:1}
\mathbf{\nabla}\cdot\mathbf{g}\,=\,-\,4\pi G \rho_0\,=\,-\,\rho_0/\epsilon_{0g},
\end{equation}
\begin{equation}
\label{eq:5:2}
\mathbf{\nabla}\times\mathbf{b} = - \frac{4\pi G}{c_g^2} \mathbf{j} + \frac{1}{c_g^2}\frac{\partial \mathbf{g} }{\partial t} = - \mu_{0g}\mathbf{j} + \frac{1}{c_g^2}\frac{\partial \mathbf{g} }{\partial t},
\end{equation} 
\begin{equation}
\label{eq:5:3}
\mathbf{\nabla}\cdot\mathbf{b}\,=\,0,
\end{equation} 
\begin{equation}
\label{eq:5:4}
\mathbf{\nabla}\times\mathbf{g}\,=\,- \frac{\partial \mathbf{b}}{\partial t}.
\end{equation} 
\end{subequations}
with $c_g$ and the gravito-Lorentz force law as stated in Eq. \eqref{eq:2:5} and Eq. \eqref{eq:3} respectively. The vector gravitational theory, represented by the Eqs. \eqref{eq:3} and \eqref{eq:5} has been named as Maxwellian Gravity (MG) by Behera and Naik \cite{14}\footnote{Who relying on McDonald's \cite{19} report of HG, stated that MG is same as HG. This should not be taken for granted without a proof because a sign difference in some vector quantities or equations has different physical meanings.} in honor of J. C. Maxwell for his first attempt in this direction. Behera and Naik \cite{14} obtained these equations from relativistic considerations, which will be revisited in this paper to obtain some new results, viz., (a) derivation the HG equations form special relativity, (b) establishment of the physical equivalence of HG and MG and (c) finding the correct gravito-Lorentz force law \eqref{eq:4} for HG. Without this correction, the effect the gravitomagnetic field of the spinning Sun on the precession of a planet's orbit has the opposite sign to the observed effect as noted in refs. \cite{19,20}. Heaviside also considered, the propagation gravitational waves carrying energy momentum in terms of gravitational analogue of electromagnetic Heaviside-Poynting's theorem.\\
Apart from Maxwell and Heaviside, prior attempts to build a relativistic theory of gravitation were based on an application of Maxwell's equations were made by Lorentz in 1900 \cite{21} and Poincar\`{e} \cite{22} in 1905. There was a good deal of debate concerning Lorentz-covariant theory of gravitation  in the years leading up to Einstein's publication of his work in 1915 \cite{5,23}. For an overview of research on gravitation from 1850 to 1915, see Roseveare \cite{24}, Renn et al. \cite{25}.  Walter \cite{26} in ref. \cite{25} discussed the Lorentz-covariant theories of gravitation. However, the success of 
Einstein's gravitation theory, described in many books \cite{4,5,18,23,24,27,28,29,30,31}, led to the abandonment of these old efforts. It seems, Einstein was unaware of Heaviside's work on gravity, otherwise his remark on Newton's theory of gravity would have been different than what he made before the 1913 congress of natural scientists in Vienna \cite{32},viz.,  
\begin{quotation}
\noindent
After the un-tenability of the theory of action at distance had thus been proved
in the domain of electrodynamics, confidence in the correctness of Newton's
action-at-a-distance theory of gravitation was shaken. One had to believe
that Newton's law of gravity could not embrace the phenomena of gravity in
their entirety, any more than Coulomb's law of electrostatics embraced the
theory of electromagnetic processes.
\end{quotation}  
However, in 1953, Sciama\cite{33} hypothetically adopted MG (by assuming gravitational mass\footnote{The mass that appears in the Newton's law of gravitostatics is called the gravitational mass, which in analogy with Coulomb's law of electrostatics may be regarded as the gravitational charge of a body.} $m_g = m_0$, the rest mass - a measure of inertia of a body at rest) to explain the origin of inertia, calling it a toy model theory of gravity which differs from general relativity (GR) principally in three respects: 
 (a) It enables the amount of matter in the universe to be estimated from a knowledge of the gravitational constant, 
 (b) The principle of equivalence is a consequence of the theory, not an initial axiom and 
 (c) It implies that gravitation must be attractive. However, he concluded his paper mentioning 
 three limitations of such a theory: (i) It is incomplete because the relativistic form of Newton's law must be derived from a tensor potential\footnote{This thought comes to anyone who believes in $m_g = E/c^2$, where $E$ is the relativistic energy (that includes the rest energy $E_0 = m_0c^2$) which may not be true as will be shown later.}, not from a vector potential, (ii) It is difficult to give a consistent relativistic discussion of the structure of the universe as a whole and (iii) It is also difficult to describe the motion of light in a gravitational field.
  Carstoiu \cite{34,35,36}, in 1969, rediscovered Heaviside's gravitational equations in the form of Eqs. \eqref{eq:5} (in our present notation as per the report of Brilloiun \cite{36}) assuming the existence of a second gravitational field called {\it gravitational vortex} (here called gravito-magnetic field) and assumed $c_g = c$ by electromagnetic analogy \cite{36}. In 1980, Cattani \cite{37} considered 
  linear equations for the gravitational field by introducing a new field by calling it the {\it Heavisidian field} which depends on the velocities of gravitational charges in the same way as a magnetic field depends on the velocities of electric charges and shown that a gravitational field may be written with linear co-variant equations in the same way as for the electromagnetic field. Cattani's equations differ from some important formulae of general relativity such as the gravitational radiation, Coriolis force by a factor of 4. In 1982, Singh \cite{15} considered a vector gravitational theory having formal symmetry with the electromagnetic theory and explained the (a) precession of the perihelion of a planet (b) bending of light in the gravitational field and (c) gravitational red-shift by postulating the self-interaction between a particle velocity and its vector potential. In 2004, Flanders and Japaridze \cite{16} axiomatically used the field equations of MG and special relativity to explain the photon deflection and perihelion advance of Mercury in the gravitational field of the Sun. Borodikhin \cite{17} explained the perihelion advance of Mercury, gravitational deflection of light as well as Shapiro time delay by postulating a vector theory of gravity in flat space-time that is nothing but MG. Borodikhin also showed that in a vector theory of gravity, there exists a model for an expanding Universe. Jefimenko \cite{13,38} also deduced the equations of MG by extending Newton's gravitational theory to time-dependent sources and fields and using the causality principle. Jefimenko assumed $c_g = c$ and postulated a gravito-Lorentz force. Recently, Heras \cite{39}, by recognizing the general validity of the 
axiomatic approach to Maxwell's equations of electromagnetic theory, used those axioms to derive only the field equations (leaving out gravito-Lorentz force law) of MG, where the 
in-variance of gravitational charge (or mass) is considered. Other recent derivations of MG equations from different approaches include the works of Nyambuya \cite{40}, Sattinger \cite{41}, Vieira and Brentan \cite{42}. The historical objections of several researchers, starting from J. C. Maxwell \cite{6} upto Misner, Thorne and Wheeler (MTW, Sec.7.2)\cite{5}, concerning negative energy density of gravitational field (`Maxwell's Enigma' as Sattinger puts it) in a linear Lorentz invariant field theory of gravity are also refuted by Sattinger \cite{41}, who considered negative field energy density for MG in agreement with the result reported by Behera and Naik \cite{5}. In the discussion on the Dark Matter problem, Sattinger further noted: ``{\it The Maxwell-Heaviside equations of gravitation constitute a linear, relativistic correction to Newton's equations of motion; they interpolate between Newton's and Einstein's theories of gravitation, and are  therefore a natural mathematical model on which to build a dynamical theory of galactic structures}". 
However, in this work the gravitational energy density for free fields is fixed positive by choice to address the objection of MTW (Sec. 7.2)\cite{5} without any inconsistency with the field equations HG or MG. A detailed discussion on the energy-momentum of gravitational field, which has a long history, is left out here for another paper. \\
In the context of General Relativity, several authors have obtained different Lorentz-Maxwell-type equations for gravity following different linearization procedures leading to different versions that are not isomorphic and have several serious limitations as seen in the recent report of Behera \cite{43}.
\section{HMG Form Special Relativity}
The fundamental equations of HG and MG will be derived here using special theory of relativity (SR), wherein we will make a correction to Heaviside's speculative gravitational analogue of the Lorentz force law and establish the physical equivalence of HG and MG. \\
With the establishment of SR and the equivalence of mass and energy, the meaning of the inertial mass and gravitational mass became ambiguous, because SR suggests two inertial mass-energy concepts: (1) the Lorentz invariant rest-mass $m_0\,=\,E_0/c^2$ ($E_0$ = rest-energy, which is the sum total of all forms of energy in the rest frame of a body or
particle) and (2) the mass attributed to the relativistic energy $m\,= \,E/c^2$ ($E\,=$
sum of all forms of energy at rest and motion) which is not Lorentz-invariant. The qualitative distinction that existed between two inertial mass concepts in Newtonian mechanics became quantitatively distinct and clear in SR. Now, one fundamental question arises, ``What form of mass (or energy) should represent the gravitational mass\footnote{ In Newtonian physics $m_g = m_0$ for Galileo's law of Universality of Free Fall to be true.} ($m_g$) in a relativistic version of Newtonian gravity?" In any construction of a field theory of gravity compatible with SR and the
correspondence principle by which a relativistic theory gravity is reducible to Newtonian gravity, a decision on which form of ``mass" (or energy) is the source of gravity has to be taken. Such a decision, as Price \cite{44} has rightly pointed out, will be
crucial not only to the resolution of the ambiguity mentioned above but also to the issue of the nonlinear nature of gravity. One of the Eddington's \cite{18} four reasons to feel dissatisfied with Newton's Law of gravitation is appropriate here to quote:
\begin{quote}
The most serious objection against the Newtonian Law as an exact law was that
it had become ambiguous. The law refers to the product of the masses of the two bodies; but the mass depends on the velocity- a fact unknown in Newton's days. Are we to take the variable mass, or the mass reduced to rest? Perhaps a learned judge, interpreting Newton's statement like a last will and testament, could give a decision; but that is scarcely the way to settle an important point in scientific theory.
\end{quote} 
In his construction of a relativistic theory of gravity popularly known as General Relativity (GR), Einstein has taken a decision in favor of the equality of $m$ with $m_g$. For a theoretical justification of this decision,
Einstein by writing Newton's equation of motion in a gravitational field (in our present mathematical notation) as
\begin{equation}
\label{eq:6}
m\frac{d\mathbf{v}}{dt}\,=\,m_g \mathbf{g}
\end{equation}
(wrongly!) inferred from it (\cite{23}, pp. 57):
\begin{quote}
It is only when there is numerical equality between the inertial and gravitational mass that the acceleration is independent of the nature of the body.
\end{quote}
This inference is often expressed in one of the two ways:\\
(S1) that the particle's motion is mass independent, or \\
(S2) that the particle's inertial mass $m$ = its
gravitational mass $m_g$.\\
The two statements (S1) and (S2) are sometimes used
interchangeably as the {\it weak equivalence principle} (WEP) in
the literature \cite{27,28,29,30,31}. This use of terminology is rather
confusing, as the two statements are logically independent \cite{45}. They happen to coincide in the context of Galileo-Newtonian physics where $m_0 = m = m_g$ but may diverge in the context of special relativity where $m \neq m_0$ and Einstein's wrong inference of $m_0\neq m = m_g$ from a non-relativistic Eq. \eqref{eq:6}, where $m = m_0$ and $m_g = m_0$ is a condition for Galileo's law of Universality of Free Fall to be true. To explore this possibility, to get new insights for making Newtonian gravity compatible with the SR, to regard old problems from a new angle, we re-examined \cite{14} an often cited
\cite{46,47} Salisbury-Menzel's \cite{48,49,50,51} thought experiment
(SMTE) from a new perspective as discussed in the following subsection containing some new thoughts and results not explicitly revealed in \cite{14}. 
Before that the authors would like to remark that perhaps Einstein, himself, was not satisfied with his above inference of $m_g = m$, as we can sense from  his another statement on the equality of $m_g$ with $m$ \cite{52,53}:
\begin{quote}
The proportionality between the inertial and gravitational masses holds for all bodies without 
exception, with the (experimental) accuracy achieved thus far, so that we may assume its general validity until proved otherwise.
\end{quote}  
The last three words of Einstein's above statement,`{\it until proved otherwise}', show that he was very cautious and not very confident of what he was stating. Based on the experimental results available up to 1993, Mashhoon \cite{47} noted that the observational evidence for the principle of equivalence of gravitational and inertial masses was not yet precise enough to reflect the wave nature of matter and radiation in their interactions with gravity (see other references on equivalence principle in \cite{14,47}).   
\subsection{Re-Examination of SMTE to Show $m_0 = m_g$}
Consider a system of two non-spinning point-like charged particles
with charges $q_1$ and $q_2$ and respective rest masses $m_{01}\, (= E_{01}/c^2)$
and $m_{02}\, (= E_{02}/c^2)$ such that they are at rest in an inertial frame
$S^\prime$ under equilibrium condition due to a mutual balance of
the force of Coulombic repulsion ($\mathbf{F}_C^\prime$) and the Newtonian 
gravitostatic attraction ($\mathbf{F}_N^\prime$) between them. Our aim is to 
investigate the condition of equilibrium of this two-particle system (realizable 
in a Laboratory by taking two perfectly identical spherical metallic spheres
having requisite masses and charges so that they are in
equilibrium) in different inertial frames in relative motion.
For our re-examination purpose, suppose that the particles are positively charged and
they are in empty space. Let the particle No.2 be positioned at
the origin of $S^\prime$-frame and $\mathbf{r}^\prime$ be the position
vector of the particle No.1 with respect to the particle No.2. In
this $S^\prime$-frame the condition of equilibrium is fulfilled by
\begin{equation}
\label{eq:7}
\mathbf{F}_C^\prime\,+\mathbf{F}_N^\prime\,=\,\frac{q_1q_2 \vec
r^\prime}{4\pi \epsilon_0 {r^\prime}^3}\,-\,\frac{Gm_{01}m_{02}\vec
r^\prime}{{r^\prime}^3} = {\bf 0},
\end{equation}
where $r^\prime\,=\,|\vec r^\prime|$ and other symbols have their usual meanings. From Eq.  \eqref{eq:7} we get
\begin{equation}
\label{eq:8}
\frac{q_1q_2}{4\pi \epsilon_0}=G
m_{01}m_{02}=\frac{m_{01}m_{02}}{4\pi \epsilon
_{0g}}\qquad(\epsilon_{0g}=1/4\pi G).
\end{equation}
Eq. \eqref{eq:8} represents the condition of equilibrium, in terms of
the charges and rest masses (or rest energies) of the particles, under which an
equilibrium can be ensured in the $S^\prime$-frame. For example,
if each metallic sphere is given a charge of $1\times 10^{-6}$
Coulomb, then the rest mass of each sphere should be $1.162\times
10^{4}$ kg, to fulfill the equilibrium condition \eqref{eq:8}
 in a laboratory experiment.  \\ 
Now, let us investigate the problem of equilibrium of the said
particle system from the point of view of an observer in another
inertial frame $S$, in uniform relative motion with respect to the
$S^\prime$-frame. To simplify the investigation, let the relative
velocity $\mathbf{v}$ of $S$ and $S^\prime$-frame be along a common
$X/X^\prime$-axis with corresponding planes parallel as usual.
Since the particles are at rest in $S^\prime$-frame, both of them
have the same uniform velocity $\mathbf{v}$ relative to the $S$-frame.
Let the position vector of the particle No.1 with respect to the
particle No.2 as observed in the $S$-frame be $\mathbf{r}$ and the angle
between $\mathbf{v}$ and $\mathbf{r}$ be $\theta$.\\
For an observer in the $S$-frame, the force of electric origin on
either particle (say on particle No.1 due to particle No.2) is no
more simply a Coulomb force, but a Lorentz force, viz.,
\begin{equation}
\label{eq:9}
\mathbf{F}_L\,=\,q_1\mathbf{E}_2\,+\,q_1\mathbf{v}\times \mathbf{B}_2
\end{equation}
where
\begin{equation}
\label{eq:10}
\mathbf{E}_2\,=\,\frac{q_2(1\,-\,\beta^2)\mathbf{r}}{4\pi\epsilon_0
r^3\left(1\,-\,\beta ^2
\sin^2{\theta}\right)^{3/2}},\qquad(\mathbf{\beta}  = \mathbf{v}/c)
\end{equation}
\begin{align}\label{eq:11}
\mathbf{B}_2\,&= \frac{\mathbf{v}\times \mathbf{E}_2}{c^2} = \frac{(q_2\mathbf{v})\times \mathbf{r}\,(1\,-\,\beta^2)}{4\pi \epsilon_0 c^2\,r^3\left(1\,-\,\beta ^2 \sin^2{\theta}\right)^{3/2}} \nonumber \\
         &= \frac{\mu_0}{4\pi}\,\frac{(q_2\mathbf{v})\times \mathbf{r}\,(1\,-\,\beta^2)}{r^3\left(1\,-\,\beta ^2
\sin^2{\theta}\right)^{3/2}}
\end{align}
\begin{equation}
\label{eq:12}
\mathbf{r}\,=\,\frac{\mathbf{r}^\prime \left(1\,-\,\beta
^2\sin^2{\theta}\right)^{1/2}}{\left(1\,-\,\beta ^2\right)^{1/2}}.
\end{equation}
What about the force of gravitational interaction as observed in
the $S$-frame? It can not simply be a Newtonian force but
something else, otherwise the particle system will not remain in
equilibrium in the $S$-frame. Such a situation will amount to a
violation of the principle of relativity in special relativity. A
null force should remain null in all inertial frames. Therefore, a
new force law of gravity has to be invoked so that the equilibrium
is maintained in accordance with the principle of relativity (Lorentz invariance of physical laws). 
Let this new unknown force be represented by $\mathbf{F}_{gL}$ such that
the equilibrium condition in $S$-frame is satisfied as:
\begin{equation}
\label{eq:13}
\mathbf{F}_{gL}\,+\,\mathbf{F}_L\,=\,\mathbf{0}\qquad\implies \, \mathbf{F}_{gL}\,= -\,\mathbf{F}_L.
\end{equation}
Taking into account the Eqs. \eqref{eq:9}-\eqref{eq:12}, $\mathbf{F}_{gL}$ in
Eq. \eqref{eq:13} can be expressed as:
\begin{align}\label{eq:14}
\vec F_{gL}\,&=\,-\,\frac{q_1q_2\left(1\,-\,\beta^2\right)\vec
r}{4\pi \epsilon _0 r^3\left(1\,-\,\beta ^2
\sin^2{\theta}\right)^{3/2}} \nonumber \\
         &\quad {} -\,\frac{\mu_0}{4\pi}\,\frac{q_1q_2\vec
v \times (\vec v\times \vec r)\left(1\,-\,\beta^2\right)}{
r^3\left(1\,-\,\beta ^2 \sin^2{\theta}\right)^{3/2}}.
\end{align}
Now, using Eq.\eqref{eq:8}, we can eliminate $q_1q_2$ from Eq. \eqref{eq:14}
to get the expression for $\vec F_{gL}$ in terms of $m_{01},
m_{02}$ and $G$ as:
\begin{align}\label{eq:15}
\vec F_{gL}\,&=\,-\,\frac{Gm_{01}m_{02}\left(1\,-\,\beta^2\right)\vec
r}{r^3\left(1\,-\,\beta ^2
\sin^2{\theta}\right)^{3/2}} \nonumber \\
         &\quad {} -\,\frac{G}{c^2}\,\frac{m_{01}m_{02}\vec
v \times (\vec v\times \vec r)\left(1\,-\,\beta^2\right)}{
r^3\left(1\,-\,\beta ^2 \sin^2{\theta}\right)^{3/2}} \nonumber \\ 
           &= \,-\,\frac{1}{4\pi \epsilon_{0g}}\frac{m_{01}m_{02}\left(1\,-\,\beta^2\right)\vec r}{r^3\left(1\,-\,\beta ^2
\sin^2{\theta}\right)^{3/2}} \nonumber \\
         &\quad {} -\,\frac{\mu_{0g}}{4\pi}\,\frac{m_{01}m_{02}\vec
v \times (\vec v\times \vec r)\left(1\,-\,\beta^2\right)}{
r^3\left(1\,-\,\beta ^2 \sin^2{\theta}\right)^{3/2}}, 
\end{align}
\noindent
where
\begin{equation}\label{eq:16}
\epsilon_{0g}\,=\,\frac{1}{4\pi G},\,\,
\mu_{0g}\,=\,\frac{4\pi G}{c^2} \implies c = \frac{1}{\sqrt{\epsilon_{0g}\mu_{0g}}}. 
\end{equation}
By comparing the quantities in Eq. \eqref{eq:16} with that of electromagnetic theory, viz., $c = (\epsilon_0\mu_0)^{-1/2}$, we immediately find that, gravitational waves, if they exist, must have a wave velocity $c_g$ in vacuum:
\begin{equation}\label{eq:17}
c_g\,=\,\frac{1}{\sqrt{\epsilon_{0g}\mu_{0g}}}\,=\,c.
\end{equation}
Now, Eq. \eqref{eq:15} may be rearranged to the following form to represent the Gravito-Lorentz force law of special relativistic Maxwellian Gravity (SRMG):
\begin{equation}\label{eq:18}
\mathbf{F}_{gL}^{MG}= m_{01}\mathbf{g}_2 +m_{01}\mathbf{v} \times \mathbf{b}_2 \qquad \text{(For MG)} 
\end{equation}
where 
\begin{align}\label{eq:19}
\mathbf{g}_2\,&= \,-\,\frac{1}{4\pi \epsilon_{0g}}\,\frac{m_{02}(1\,-\,\beta^2)\mathbf{r}}{r^3\left(1\,-\,\beta^2\sin^2\theta\right)^{3/2}} \nonumber \\
          &\simeq  -\,\frac{1}{4\pi \epsilon_{0g}}\,\frac{m_{02}\mathbf{r}}{r^3} 
          \qquad (\mbox{when}\, \beta << 1),
\end{align}
\begin{align}\label{eq:20}
\mathbf{b}_2\,&= \frac{\mathbf{v} \times \mathbf{g}_2}{c^2}\,=\,-\,\frac{\mu_{0g}}{4\pi}\,\frac{(m_{02}\mathbf{v})\times \mathbf{r}\,(1\,-\,\beta^2)}{r^3\left(1\,-\,\beta^2\sin^2\theta\right)^{3/2}} \nonumber \\
          &\simeq  -\,\frac{\mu_{0g}}{4\pi}\,\frac{(m_{02}\mathbf{v})\times \mathbf{r}}{r^3} \qquad (\mbox{when}\, \beta << 1).
\end{align}
Eqs. \eqref{eq:18}-\eqref{eq:20} are in complete formal analogy with the Eqs. \eqref{eq:9}-\eqref{eq:11} of classical electromagnetism in its relativistic version. Thus, from the requirement of the frame-independence of the equilibrium condition, we not only obtained a gravitational analogue of the Lorentz-force law expressed by Eq. \eqref{eq:18} but also unexpectedly found the Lorentz-invariant rest mass as the gravitational analogue of the electric charge by electromagnetic analogy. From this analysis, the gravitational charge (or rest mass) invariance may be interpreted as a consequence of the Lorentz-invariance of the physical laws. These findings are in conformity with Poinca\`{r}e's \cite{54} remark that 
{\it if equilibrium is to be a frame-independent condition, it is necessary for all forces of non-electromagnetic origin to have precisely the same transformation law as that of the Lorentz-force}. 
\noindent
Now, following Rosser's \cite{55} approach to classical electromagnetism via relativity,  
Behera and Naik \cite{14} and Behera \cite{38} had obtained the field equations of MG as given by the Eqs. \eqref{eq:5}. \\
On the other hand, if one retains the definition of $\mathbf{g}_2$ as in Eq. \eqref{eq:19} and redefines $\mathbf{b}_2$ as
\begin{align}\label{eq:21}
\mathbf{b}_2\,&=\,-\,\frac{\mathbf{v} \times \mathbf{g}_2}{c^2}\,=\,\frac{\mu_{0g}}{4\pi}\,\frac{(m_{02}\mathbf{v})\times \mathbf{r}\,(1\,-\,\beta^2)}{r^3\left(1\,-\,\beta^2\sin^2\theta\right)^{3/2}} \nonumber \\
          &\simeq  \,\frac{\mu_{0g}}{4\pi}\,\frac{(m_{02}\mathbf{v})\times \mathbf{r}}{r^3} \qquad (\mbox{when}\, \beta << 1),
\end{align}
then the gravito-Lorentz force law \eqref{eq:18} must be of the form as in Eq. \eqref{eq:4}
so as to describe the same physics (or physical effects) as implied by Eqs. \eqref{eq:18}-\eqref{eq:20} or their source Eq. \eqref{eq:15}. The field equations that are consistent with the equations \eqref{eq:19}, \eqref{eq:21} and \eqref{eq:4} can again be obtained following Rosser \cite{55}. These represent Heaviside's gravitational field equations as originally proposed by him  \cite{7,9,10,11,12,13}, now written in our present notation and convention as in the Eqs. \eqref{eq:2}.
The equation of continuity 
\begin{equation}
\label{eq:22}
\mathbf{\nabla}\cdot\mathbf{j} + \frac{\partial \rho_0}{\partial t} = 0
\end{equation}
follows from the in-homogeneous equations of HG and MG. In vacuum (where $\rho_0 = 0,\mathbf{j} = \mathbf{0}$), the field equations of HG and MG give us the wave equations for the $\mathbf{g}$ and $\mathbf{b}$ fields: 
\begin{subequations}\label{eq:23}
\begin{align}
\mathbf{\nabla}^2\mathbf{g} = \frac{1}{c^2}\frac{\partial ^2 \mathbf{g}}{\partial t^2} \label{eq:23:1} \\
\mathbf{\nabla}^2\mathbf{b} = \frac{1}{c^2}\frac{\partial ^2 \mathbf{b}}{\partial t^2}\label{eq:23:2}
\end{align} 
\end{subequations}
which show that the wave velocity of gravitational waves in vacuum $c_g = c$.\\
Alternatively, after recognizing our new findings from the above thought experiment, especially $m_g = m_0$ and $c_g = c$ from Eq. \eqref{eq:17}, one may follow the following procedure to arrive at the field equations of MG and HG. 
\subsubsection{Alternative derivation of Field Equations of MG and HG}
We take for granted the Gauss's law of gravitostatics \eqref{eq:2:1} and the equation of continuity \eqref{eq:22} as valid laws of physics. To establish a link between the Eqs. \eqref{eq:2:1} and \eqref{eq:22}, we take the time derivative of Eq. \eqref{eq:2:1} and write the result as 
\begin{equation}\label{eq:24}
\frac{\partial \rho_0}{\partial t} = - \frac{1}{4\pi G}\mathbf{\nabla}\cdot \left(\frac{\partial \mathbf{g}}{\partial t}\right)
\end{equation}  
From the equation of continuity \eqref{eq:22} and the Eq. \eqref{eq:24}, we get 
\begin{equation}\label{eq:25}
\mathbf{\nabla}\cdot \left(\mathbf{j} - \frac{1}{4\pi G}\frac{\partial \mathbf{g}}{\partial t}\right) 
= \mathbf{\nabla}\cdot \left(\mathbf{j} - \epsilon_{0g}\frac{\partial \mathbf{g}}{\partial t}\right)=  0. 
\end{equation}
Now we multiply Eq. \eqref{eq:25} by $\mu_{0g} = 4\pi G/c^2$, as defined in Eq. \eqref{eq:17}, to obtain the equation: 
\begin{equation}\label{eq:26}
\mathbf{\nabla}\cdot \left(\mu_{0g}\mathbf{j} - \frac{1}{c^2}\frac{\partial \mathbf{g}}{\partial t}\right)=  0. 
\end{equation}
The quantity inside the parenthesis of Eq. \eqref{eq:26} is a vector whose divergence is zero. Since $\mathbf{\nabla}\cdot(\mathbf{\nabla}\times\mathbf{X})\,=\,0$\, for any vector $\mathbf{X}$, the vector inside the parenthesis of Eq. \eqref{eq:26} can be expressed as the curl of some other vector, say $\mathbf{b}$. Mathematically speaking, the Eq. \eqref{eq:26} admits of two independent solutions:
\begin{equation}\label{eq:27}
\mathbf{\nabla}\times \mathbf{b}\,=\,\begin{cases} 
+\mu_{0g}\mathbf{j} - \frac{1}{c^2}\frac{\partial \mathbf{g}}{\partial t} & \quad   
\text{(For HG)}  \\
-\mu_{0g}\mathbf{j} + \frac{1}{c^2}\frac{\partial \mathbf{g}}{\partial t} & \quad  
\text{(For MG)}  
\end{cases}
\end{equation}
Thus, we arrived at the Eq. \eqref{eq:2:2} of HG and Eq. \eqref{eq:5:2} of MG. 
In vacuum ($\mathbf{j} = \mathbf{0}$), the Eqs. \eqref{eq:27} become
\begin{equation}\label{eq:28}
\mathbf{\nabla}\times \mathbf{b}\,=\,\begin{cases} 
- \frac{1}{c^2}\frac{\partial \mathbf{g}}{\partial t} & \quad   \text{(For HG)} \\
+ \frac{1}{c^2}\frac{\partial \mathbf{g}}{\partial t} & \quad   \text{(For MG)}  
\end{cases}
\end{equation} 
Taking the curl of the Eqs. \eqref{eq:28} we get 
\begin{equation}\label{eq:29}
\mathbf{\nabla}(\mathbf{\nabla}\cdot \mathbf{b})- \mathbf{\nabla}^2\mathbf{b}\,=\,
\begin{cases} 
- \frac{1}{c^2}\frac{\partial}{\partial t}(\mathbf{\nabla}\times \mathbf{g}) & \quad 
\text{(For HG)} \\ 
+ \frac{1}{c^2}\frac{\partial}{\partial t}(\mathbf{\nabla}\times \mathbf{g}) & \quad 
 \text{(For MG)} 
\end{cases}
\end{equation}
The Eqs. \eqref{eq:29} will reduce to the wave equation \eqref{eq:23:2} for the $\mathbf{b}$ field, if the following conditions: 
\begin{equation}\label{eq:30}
\mathbf{\nabla}\cdot \mathbf{b} = 0 \qquad    \text{(For both HG and MG)} 
\end{equation}
\begin{equation}\label{eq:31}
\mathbf{\nabla}\times \mathbf{g}\,=\,\begin{cases} 
+ \frac{\partial \mathbf{b}}{\partial t} & \quad   \text{(For HG)} \\ 
- \frac{\partial \mathbf{b}}{\partial t} & \quad   \text{(For MG)} 
\end{cases}
\end{equation}
are satisfied. Thus, we arrived at the Eqs. \eqref{eq:2:3}-\eqref{eq:2:4} of HG and Eqs. \eqref{eq:5:3}-\eqref{eq:5:4} of MG by imposing the condition of existence of gravitational waves in vacuum. This way we found the correctness of the original gravitational field equations found by Heaviside and as seen in \cite{7,9,10,11,12,13} and corrected Heaviside's gravito-Lorentz force law to the form as given in Eq. \eqref{eq:4} so as to be consistent with his field equations \eqref{eq:2:1}-\eqref{eq:2:4}). It is due an error in the sign in the gravitomagnetic force term that the effect of gravitomagnetic filed of the spinning Sun on the precession of a planet's orbit has the opposite sign to the observed effect as noted by McDonald \cite{19} and Iorio \cite{20}, who did not trace the cause of this error. As per our present relativistic study, we found HG and MG to represent the same physical phenomena, the sign differences in some terms in their equations are attributed to the definitions of some physical quantities. Thus, HG and MG are mere two mathematical representations of a single vector theory of gravity named here as Heaviside-Maxwellian Gravity (HMG).  \\
\indent
Since $\mathbf{\nabla}\cdot \mathbf{b}\,=\,0$ for both HG and MG, $\mathbf{b}$ can be defined as the curl of some vector function, say $\mathbf{A}_g$. If we define, 
\begin{equation}\label{eq:32}
\mathbf{b}\,=\,\begin{cases} 
-\mathbf{\nabla}\times \mathbf{A}_g & \quad   \text{(For HG)} \\ 
+\mathbf{\nabla}\times \mathbf{A}_g & \quad   \text{(For MG)} 
\end{cases}
\end{equation}
then using these definitions in Eqs. \eqref{eq:31}, we find 
\begin{equation}\label{eq:33}
\mathbf{\nabla}\times \left(\mathbf{g} + \frac{\partial \mathbf{A}_g}{\partial t}\right)\,=\,\mathbf{0} \quad \text{(For both MG and HG)},
\end{equation}
which is equivalent to say that the vector quantity inside the parentheses of Eq. \eqref{eq:33} can be written as the gradient of a scalar potential, $\phi_g$: 
\begin{equation}\label{eq:34}
\mathbf{g}\,=\,-\,\mathbf{\nabla}\phi_g \,-\,\frac{\partial \mathbf{A}_g}{\partial t} \quad \text{(For both MG and HG)}.
\end{equation}
Substituting the expression for $\mathbf{g}$ given by Eq. \eqref{eq:34} and the expression for $\mathbf{b}$ defined by Eq. \eqref{eq:32} in the in-homogeneous field Eqs. \eqref{eq:2:1}-\eqref{eq:2:2} of HG and \eqref{eq:5:1}-\eqref{eq:5:2} of MG, we get the following expressions for their in-homogeneous equations in terms of scalar and vector potentials as
\begin{equation}\label{eq:35}
\mathbf{\nabla}^2\phi_g - \frac{1}{c^2} \frac{\partial ^2\phi_g}{\partial t^2}= \frac{\rho_0}{\epsilon_{0g}} \quad \text{(For both MG and HG)}, 
\end{equation}
\begin{equation}\label{eq:36}
\mathbf{\nabla}^2\mathbf{A}_g - \frac{1}{c^2} \frac{\partial ^2\mathbf{A}_g}{\partial t^2}
=\mu_{0g}\mathbf{j} \quad \text{(For both MG and HG)}, 
\end{equation}
\noindent
if the following gravitational Lorenz gauge condition, 
\begin{equation}\label{eq:37}
\mathbf{\nabla}\cdot\mathbf{A}_g\,+\,\frac{1}{c^2}\frac{\partial \phi_g}{\partial t}\,=\,0 \quad \text{(For both MG and HG)},
\end{equation}
is imposed. These will determine the generation of gravitational waves by prescribed gravitational mass and mass current distributions. Particular solutions of Eq. \eqref{eq:35} and Eq. \eqref{eq:36} in vacuum are
\begin{equation}\label{eq:38}
\phi_g(\mathbf{r}, t) = -\frac{1}{4\pi \epsilon_{0g}}\int \frac{\rho_0(\mathbf{r}^\prime , t^\prime)}{|\mathbf{r} -\mathbf{r}^\prime|}dv^\prime \quad {\mbox{and}}
\end{equation}
\begin{equation}\label{eq:39}
\mathbf{A}_g(\mathbf{r}, t) = -\frac{\mu_{0g}}{4\pi }\int \frac{\mathbf{j}(\mathbf{r}^\prime , t^\prime)}{|\mathbf{r} -\mathbf{r}^\prime|}dv^\prime,
\end{equation}
\noindent
where $t^\prime = t - |\mathbf{r} - \mathbf{r}^\prime|/c$ is the retarded time and $dv^\prime$ is an elementary volume element at $\mathbf{r}^\prime$. Thus, we saw that retardation in gravity is possible in flat space-time in the same procedure as we adopt in  electrodynamics. Hence, we have reasons to strongly disagree with those who believe in Rohrlich's conclusion \cite{56}: ``{\it Because the Newtonian theory is entirely static, retardation is not possible until the correction due to deviations from Minkowski space is considered"}.  Before passing to the next section, the authors wish to note that using our present approach to the discovery of HG and MG, one can obtain a new mathematical form of Lorentz-Maxwell's equations, which are physically equivalent to the standard Lorentz-Maxwell's equations. These new form of Lorentz-Maxwell's equations are noted in the following box.
\begin{tcolorbox} 
{\bf New Form of Lorentz-Maxwell Equations of Electrodynamics:} \\
\begin{eqnarray*}
\mathbf{\nabla}\cdot\mathbf{E}\,=\,\rho_e/\epsilon_{0},  \\
\mathbf{\nabla}\times\mathbf{B}_{\text{new}} = \,-\,\mu_{0}\mathbf{j}_e - \frac{1}{c^2}\frac{\partial \mathbf{E} }{\partial t},\\
\mathbf{\nabla}\cdot\mathbf{B}_{\text{new}}\,=\,0,\\
\mathbf{\nabla}\times\mathbf{E}\,=\,\frac{\partial \mathbf{B}_{\text{new}}}{\partial t},
\end{eqnarray*}
where 
\begin{equation*}
 c = \frac{1}{\sqrt{\epsilon_{0}\mu_{0}}}
\end{equation*}
\begin{equation*}
\mathbf{F}_L^{\text{new}} = q\left(\mathbf{E}\,-\,\mathbf{v}\times \mathbf{B}_{\text{new}}\right)
\end{equation*}
\begin{equation*}
 \mathbf{B}_{\text{new}} = - \mathbf{\nabla}\times \mathbf{A}_{\text{new}} 
\end{equation*}
\begin{equation*}
 \mathbf{E} = - \mathbf{\nabla}\phi_e - \frac{\partial \mathbf{A}_{\text{new}} }{\partial t}  
\end{equation*}
\end{tcolorbox}
\subsection{Lorentz co-variant formulation of MG} 
In the Lorentz co-variant formulation, by introducing the space-time 4-vector $x^\alpha = (ct, \vec{x})$, proper (or rest) mass current density 4-vector $j^\alpha = (\rho_o c,\, \mathbf{j})$, $j_\alpha = (\rho_oc,\,- \mathbf{j})$, $A_{g\alpha} = (\phi_g/c,\,- \mathbf{A}_g)$ and $A_g^\alpha = (\phi_g/c,\,\mathbf{A}_g)$; $\partial_\alpha \equiv \left(\partial/{c\partial t},\,\mathbf{\nabla}\right) \& \, \partial^\alpha \equiv \left(\partial/{c\partial t},\,-\mathbf{\nabla}\right)$ and second-rank anti-symmetric gravitational field strength tensor $f_{\alpha \beta}$ for MG\footnote{Here we left out the case for HG, which the reader may try.}
\begin{equation}\label{eq:40}
f_{\alpha \beta}\,= \partial_\alpha A_{g\beta} - \partial_\beta A_{g\alpha} =
\begin{pmatrix}
  0         &    \frac{g_{x}}{c}    &    \frac{g_{y}}{c}   &   \frac{g_{z}}{c}  \\
-\frac{g_{x}}{c}   &    0       &   -b_{z}         &   b_{y}  \\ 
-\frac{g_{y}}{c}   &    b_{z}  &    0              &  -b_{x}  \\ 
-\frac{g_{z}}{c}   &   -b_{y}  &    b_{x}         &   0       
\end{pmatrix},
\end{equation}
one can rewrite the field equations of MG as:
\begin{equation}\label{eq:41}
\partial^\beta f_{\alpha \beta}\,= \partial^\beta\left(\partial_\alpha A_{g\beta} - \partial_\beta A_{g\alpha}\right)=\,\frac{4\pi G}{c^2}j_\alpha = \,\mu_{0g}j_\alpha,
\end{equation}
\begin{equation}\label{eq:42}
\partial_\alpha f_{\beta \gamma} +\partial_\beta f_{\gamma \delta} + \partial_\gamma f_{\alpha \beta} = 0,
\end{equation}
where $\alpha,\, \beta,\,\gamma$ are any three of the integers 0, 1, 2, 3;  and the Gravito-Lorenz condition: $\partial^\alpha A_{g\alpha} = 0$.  Now, the relativistic gravito-Lorentz force law of MG takes the following form 
\begin{equation}\label{eq:43}
\frac{d^2x^\alpha}{d\tau^2} = f^{\alpha \beta}\frac{dx_\beta}{d\tau},
\end{equation}
where $\tau$ is the proper time along the particle's world-line and $f^{\alpha \beta}$ is given by 
\begin{equation}\label{eq:44}
f^{\alpha \beta} =\eta^{\alpha \gamma}f_{\gamma \delta}\eta^{\delta \beta} = 
\begin{pmatrix}
  0              & -\frac{g_{x}}{c} &   -\frac{g_{y}}{c}   &  -\frac{g_{z}}{c} \\
\frac{g_{x}}{c} & 0                 &   -b_{z}             &   b_{y}  \\ 
\frac{g_{y}}{c} & b_{z}            &    0                  &  -b_{x}  \\ 
\frac{g_{z}}{c} & -b_{y}           &    b_{x}             &   0
\end{pmatrix},
\end{equation}
where the flat space-time metric tensor $\eta_{\alpha \beta} = \eta^{\alpha \beta}$ is represented by symmetric diagonal matrix with  
\begin{equation}\label{eq:45}
\eta_{00} = 1,\quad \eta_{11} = \eta_{22} = \eta_{33} = -1. 
\end{equation}
The relativistic equation of motion \eqref{eq:43} is independent of the mass of the particle moving in an external gravito-electromagnetic (GEM) field $f^{\alpha \beta}$. Thus we saw that the motion of a particle in an external GEM field can be independent of its mass without any postulation on the equality of gravitational mass with frame-dependent inertial mass. Equation \eqref{eq:43} is the relativistic generalization of Galileo's law of Universality of Free Fall (UFF) expressed through the non-relativistic equations of motion \eqref{eq:6} and known to be true both theoretically and experimentally since Galileo's time.  \\
Now, if we introduce the energy momentum four vector:
\begin{equation}\label{eq:46}
p^\alpha = (p_0,\,\mathbf{p}) = m_0(U_0,\,\mathbf{U})
\end{equation}
where $p_0 = E/c$ and $U^\alpha = \left(\gamma c, \,\gamma \mathbf{v} \right)$ is the 4-velocity, $\gamma = \left(1 - v^2/c^2\right)^{-1/2}$ is the Lorentz factor, then with this $p^\alpha$ we can re-write Eq. \eqref{eq:43} as 
\begin{equation}\label{eq:47}
\frac{dp^\alpha}{d\tau} = f^{\alpha \beta}p_\beta.
\end{equation}
Thus, the fields $f_{\alpha \beta}$ of MG couple to the energy-momentum 4-vector of all particles of whatever rest masses they have, provided $m_g = m_0$ holds exactly. It is to be noted that the equation of motion  \eqref{eq:43} holds only in an inertial frame. Appropriate modifications are necessary for its application in non-inertial frames, as is done in non-relativistic physics by introducing pseudo-forces.\\
One can verify that the equations of motion of the fields of MG can be obtained using the Euler-Lagrange equations of motion:
\begin{equation}\label{eq:48}
\partial^\beta\frac{\partial \mathcal{L}_{\text{MG}}}{\partial(\partial^\beta A_g^\alpha)}\,=\,\frac{\partial \mathcal{L}_{\text{MG}} }{\partial A_g^\alpha},
\end{equation}
where the Lagrangian density for MG is chosen as
\begin{equation}\label{eq:49}
\mathcal{L}_{\text{MG}}\,=\,-\,\frac{c^2}{16\pi G }f^{\mu \nu}f_{\mu \nu}\,+\,j^\mu A_{g\mu}.
\end{equation}
The negative sign before the first term on the right hand side of Eq. \eqref{eq:49} is fixed by choice to fulfill our requirement that the corresponding free Hamiltonian (or, better, energy densities) be positive and definite. 

\subsection{Original analysis of SMTE with assumption of $m_g = m = m_0/\sqrt{1 - v^2/c^2}$}
In the original analysis of SMTE \cite{48} Salisbury and Menzel (SM) axiomatically used flat space-time and assumed $m_g = m = m_0/\sqrt{1 - v^2/c^2}$ for their thought experimental demonstration of gravito-magnetic field (they called it {\it Gyron field}) and the gravitational analogue of Lorentz force law. From the analysis of their results, one can find that in the slow motion approximation, if the gravito-Lorentz force law is written in the following form
\begin{equation}\label{eq:50}
\mathbf{F}_{gL}^{SM} = m_0\frac{d\mathbf{v}}{dt} = m_0 \mathbf{g} + m_0\mathbf{v}\times \mathbf{b}, \quad \text{then} 
\end{equation} 
\begin{equation}\label{eq:51}
\mu_{0g}^{SM} = \frac{8\pi G}{c^2} \quad \text{while} \quad \epsilon_{0g}^{SM} = \frac{1}{4\pi G},
\end{equation}
which yields 
\begin{equation}\label{eq:52}
c_g^{SM} = \left(\mu_{0g}^{SM}\mu_{0g}^{SM}\right)^{-1/2} = c/\sqrt{2}.
\end{equation} 
On the other hand if one considers $c_g^{SM} = c$, then Eq. \eqref{eq:50} has to be written in the following form: 
\begin{equation}\label{eq:53}
\mathbf{F}_{gL}^{SM} = m_0\frac{d\mathbf{v}}{dt} = m_0 \mathbf{g} + 2m_0\mathbf{v}\times \mathbf{b}. 
\end{equation}
We designate this type of gravity as linearized version of non-linear special relativistic MG (SRGM-N) in flat space-time. The origins of the non-linearity of SRGM-N, the appearance of the spurious value of $c_g = c/\sqrt{2}$ or a factor of ``2" in the gravitomagnetic force term (due to a supposed value of $c_g = c$) are all now traced to the adoption of Einstein's doubtful postulate on the equality of gravitational mass with the velocity dependent inertial mass. 
\section{HMG Form Local Gauge Invariance}
It is well known that the free Dirac Lagrangian density\footnote{We adopt SI units in this paper for clarity to the general reader.}
\begin{equation}\label{eq:54}
\mathcal{L} = i\hbar c\overline{\psi}\gamma^\mu\partial_\mu \psi - m_0c^2\overline{\psi}\psi \qquad \text{(in SI units)}
\end{equation}
is invariant under the transformation 
\begin{equation}\label{eq:55}
\psi \rightarrow e^{i\theta}\psi \qquad {(\mbox{global phase transformation})}
\end{equation}
where $\theta$ is any real number. This is because under global phase transformation \eqref{eq:55} $\overline{\psi} \rightarrow e^{-i\theta}\overline{\psi}$ which leaves $\overline{\psi}\psi$ in Eq. \eqref{eq:54} unchanged as the exponential factors cancel out. But the Lagrangian density \eqref{eq:54} is not invariant under the following transformation 
\begin{equation}\label{eq:56}
\psi \rightarrow e^{i\theta(x)}\psi \qquad {(\mbox{local phase transformation})}
\end{equation} 
where $\theta$ is now a function of space-time $x=x^\mu=(ct, \mathbf{x})$, because the factor $\partial_\mu \psi$ in \eqref{eq:54} now picks up an extra term from the derivative of $\theta(x)$: 
\begin{equation}\label{eq:57}
\partial_\mu \psi \rightarrow \partial_\mu\left(e^{i\theta(x)}\psi\right)= i\left(\partial_\mu\theta\right)e^{i\theta}\psi + e^{i\theta}\partial_\mu \psi
\end{equation}
so that under local phase transformation,
\begin{equation}\label{eq:58}
\mathcal{L} \rightarrow \mathcal{L}^\prime = \mathcal{L} - \hbar c \left(\partial_\mu\theta\right)\overline{\psi}\gamma^\mu\psi.
\end{equation}
For massive particles ($m_0 \neq 0$), we can re-write the transformed Lagrangian density $\mathcal{L}^\prime$ in Eq. \eqref{eq:58} as
\begin{equation}\label{eq:59}
\begin{split}
\mathcal{L}^\prime &= \mathcal{L} - \hbar c \left(\partial_\mu\theta\right)\overline{\psi}\gamma^\mu\psi \\
                   &= \mathcal{L} - \left[\partial_\mu \left(\frac{\hbar}{m_0}\theta \right) \right]
                   m_0c\overline{\psi}\gamma^\mu\psi = \mathcal{L} - j^\mu \partial_\mu \lambda (x) 
\end{split}
\end{equation}
where 
\begin{equation}\label{eq:60}
j^\mu = m_0c(\overline{\psi}\gamma^\mu\psi) = \text{4-current momentum density}, 
\end{equation}
and $\lambda (x)$ stands for 
\begin{equation}\label{eq:61}
\lambda (x) = \frac{\hbar }{m_0}\theta (x) = \frac{\hbar c}{m_0c}\theta (x). 
\end{equation}
In terms of $\lambda$, then, 
\begin{equation}\label{eq:62}
\mathcal{L} \rightarrow \mathcal{L}^\prime = \mathcal{L} - j^\mu\partial_\mu \lambda,
\end{equation}
under the local transformation 
\begin{equation}\label{eq:63}
\psi \rightarrow e^{im_0\lambda(x)/\hbar }\psi. 
\end{equation}
Now, {\it we demand that the complete Lagrangian be invariant under local phase transformations}. Since, the free Dirac Lagrangian density \eqref{eq:54} is not locally phase invariant, we are forced to add something to swallow up or nullify the extra term in Eq. \eqref{eq:62}. Specifically, we suppose 
\begin{equation}\label{eq:64}
\mathcal{L} = [i\hbar c\overline{\psi}\gamma^\mu\partial_\mu \psi - m_0c^2\overline{\psi}\psi]\,
+\,j^\mu A_{g\mu},
\end{equation} 
where $A_{g\mu}$ is some new field, which changes (in coordination with the local phase transformation of $\psi$ according to the rule 
\begin{equation}\label{eq:65}
A_{g\mu} \rightarrow A_{g\mu} + \partial_\mu\lambda.
\end{equation}
This `new, improved' Lagrangian is now locally invariant. But this was ensured at the cost of introducing a new vector field that couples to $\psi$ through the last term in Eq. \eqref{eq:64}. But the Eq. \eqref{eq:64} is devoid of a `free' term for the field $A_{g\mu}$ (having the dimensions of velocity: $[L][T]^{-1}$) itself. Since it is a vector, we look to the Proca-type Lagrangian \cite{57}:
\begin{equation}\label{eq:66}
\mathcal{L}_{\mbox{free}} = \,-\,\frac{\kappa}{4}f^{\mu \nu}f_{\mu \nu} + \kappa_0\left(\frac{m_ac}{\hbar}\right)^2A_g^\mu A_{g\mu}
\end{equation}
where $\kappa > 0$ (positive) and $\kappa_0$ are some dimensional constants and $m_a$ is the mass of the free field $A_{g\mu}$. But there is a problem here, for whereas 
\begin{equation}\label{eq:67}
f^{\mu \nu} = (\partial^\mu A_g^\nu - \partial ^\nu A_g^\mu )\quad \text{or}\quad f_{\mu \nu} = (\partial_\mu A_{g\nu} - \partial _\nu A_{g\mu}) 
\end{equation}
is invariant under \eqref{eq:65}, $A_g^\mu A_{g\mu}$ is not. {\it Evidently, the new field must be mass-less} ($m_a = 0$), otherwise the invariance will be lost. The negative sign before $\kappa$ in Eq. \eqref{eq:66} is fixed by choice to fulfill our requirement that the corresponding free Hamiltonian (or, better, energy densities) be positive and definite. The complete Lagrangian density then becomes
\begin{equation}\label{eq:68}
\mathcal{L} = [i\hbar c\overline{\psi}\gamma^\mu\partial_\mu \psi - m_0c^2\overline{\psi}\psi]\,
+\,\mathcal{L}_{\mbox{new}}
\end{equation}
where 
\begin{equation}\label{eq:69}
\mathcal{L}_{\mbox{new}}\,=\,-\,\frac{\kappa}{4}f^{\mu \nu}f_{\mu \nu}\,+\,j^\mu A_{g\mu}.
\end{equation}
 The equation of motion of this new field can be obtained using the Euler-Lagrange equations of motion:
\begin{equation}\label{eq:70}
\partial^\beta\frac{\partial \mathcal{L}_{\mbox{new}}}{\partial(\partial^\beta A_g^\alpha)}\,=\,\frac{\partial \mathcal{L}_{\mbox{new}} }{\partial A_g^\alpha}.
\end{equation}
A bit calculation (see for example, Jackson \cite{57}) yields 
\begin{equation}\label{eq:71}
\frac{\partial \mathcal{L}_{\mbox{new}}}{\partial(\partial^\beta A_g^\alpha)}
\,=\,\kappa f_{\alpha \beta},
\end{equation}
and 
\begin{equation}\label{eq:72}
\frac{\partial \mathcal{L}_{\mbox{new}} }{\partial A_g^\alpha}\,=\,j_\alpha.
\end{equation}
Using Eqs. \eqref{eq:71} and \eqref{eq:72} in Euler-Lagrange Eq. \eqref{eq:70}, we get the equations of motion of the new field as 
\begin{equation}\label{eq:73}
\partial^\beta f_{\alpha \beta}\,=\,\frac{1}{\kappa}j_\alpha.
\end{equation}
Eqs. \eqref{eq:73} express the generation of $f_{\alpha \beta}$ fields by the 4-current momentum density associated with the proper (or rest) mass of neutral massive Dirac particles. However, for classical fields, the 4-current momentum density is represented by 
\begin{equation}\label{eq:74}
j^\alpha = (c\rho_0,\, \mathbf{j}), \quad \qquad   j_\alpha = (c\rho_0,\, -\mathbf{j})
\end{equation}
where $\mathbf{j} = \rho_0\mathbf{v}$, with $\rho_0 = $ proper mass density.
For static mass distributions, the current density $j_\alpha = j_0 = c\rho_0$. It produces a time-independent - static - field, given by Eqs. \eqref{eq:73}:
\begin{equation}\label{eq:75}
\cancelto{0}{\frac{1}{c}\frac{\partial f_{00}}{\partial t}} -\frac{\partial f_{01}}{\partial x} - \frac{\partial f_{02}}{\partial y} - \frac{\partial f_{03}}{\partial z} = \frac{\rho_0c}{\kappa}
\end{equation} 
where we use $\left[\partial_\alpha \equiv \left(\partial/{c\partial t},\,\mathbf{\nabla}\right) \& \, \partial^\alpha \equiv \left(\partial/{c\partial t},\,-\mathbf{\nabla}\right) \right]$. Multiplying Eq. \eqref{eq:75} by $c$ we get 
\begin{equation}\label{eq:76}
\frac{\partial (cf_{01})}{\partial x} + \frac{\partial (cf_{02})}{\partial y} + \frac{\partial (cf_{03})}{\partial z} = - \frac{\rho_0c^2}{\kappa}.
\end{equation} 
Eq. \eqref{eq:76} gives us Newton's gravitational field ($\mathbf{g}$) as expressed in the Gauss's law of gravitostatics \eqref{eq:5:1}, viz., 
\begin{equation}\label{eq:77}
\mathbf{\nabla}\cdot\mathbf{g} = \frac{\partial g_{x}}{\partial x}+\frac{\partial g_{y}}{\partial y} + \frac{\partial g_{z}}{\partial z} =\,-\, 4 \pi G \rho_0,  
\end{equation}
if we make the following identifications:
\begin{equation}\label{eq:78}
f_{01} = \frac{g_x}{c},\,\,f_{02} = \frac{g_y}{c},\,\,f_{03} = \frac{g_z}{c}\,\, \text{and} \,\,\kappa = \frac{c^2}{4 \pi G}.
\end{equation}
With these findings, we write Eq. \eqref{eq:73} as 
\begin{equation}\label{eq:79}
\partial^\beta f_{\alpha \beta}\,=\,\frac{4\pi G}{c^2}j_\alpha,
\end{equation}
which is applicable to Dirac current density \eqref{eq:60} as well as classical current density \eqref{eq:74}.
From the anti-symmetry property of $f^{\alpha \beta}$ ($f^{\alpha \beta}\,=\,-\,f^{\beta \alpha}$), it follows form the results \eqref{eq:78} that
\begin{equation}\label{eq:80}
f_{10} = - \frac{g_x}{c},\,\, f_{20}= - \frac{g_y}{c},\,\,f_{30} = - \frac{g_z}{c}\, \text{and}\,\,f_{\alpha \alpha} = 0. 
\end{equation}
The other elements of $f_{\alpha \beta}$ can be obtained as follows. For $\alpha = 1$, i.e.  $j_1 = - j_x$, Eq. \eqref{eq:79} gives us 
\begin{equation}\label{eq:81}
\begin{split}
- \frac{4\pi G}{c^2}j_x &= \frac{4\pi G}{c^2}j_1 \\
                      &= \partial^0f_{10} +\cancelto{0}{\partial ^1f_{11}}+\partial^2f_{12} + \partial ^3f_{13} \\ 
                      &= -\frac{1}{c^2}\frac{\partial g_x}{\partial t} -\frac{\partial f_{12}}{\partial y} - \frac{\partial f_{13}}{\partial z} \\
                      &= \begin{cases}
                       -\frac{1}{c^2}\frac{\partial g_x}{\partial t} + \left(\mathbf{\nabla}\times\mathbf{b}\right)_x & \quad \text{(For MG)}   \\
                      -\frac{1}{c^2}\frac{\partial g_x}{\partial t} - \left(\mathbf{\nabla}\times\mathbf{b}\right)_x & \quad\text{(For HG)}
                      \end{cases}
\end{split}
\end{equation}
where $f_{12} = - b_z$ and $f_{13} = b_y$ for Maxwellian Gravity (MG); $f_{12} = b_z$ and $f_{13} = - b_y$ for Heaviside Gravity (HG). This way, we determined all the elements of 
the anti-symmetric `field strength tensor' $f_{\alpha \beta}$:  
\begin{equation}\label{eq:82}
f_{\alpha \beta}\,=\,\begin{cases}
\begin{pmatrix}
  0      &    \frac{g_x}{c}   &   \frac{g_y}{c}   &   \frac{g_z}{c} \\
-\frac{g_x}{c}   &    0      &   -b_z     &   b_y  \\ 
-\frac{g_y}{c}   &    b_z    &    0       &  -b_x  \\ 
-\frac{g_z}{c}   &   -b_y    &    b_x     &   0       
\end{pmatrix}
& \quad \text{(For MG)}   \\ \\
\begin{pmatrix}
  0      &    \frac{g_x}{c}   &   \frac{g_y}{c}   &   \frac{g_z}{c}\\
-\frac{g_x}{c}   &    0      &    b_z     &  -b_y  \\ 
-\frac{g_y}{c}  &   -b_z    &    0       &   b_x  \\ 
-\frac{g_z}{c}   &    b_y    &   -b_x     &   0       
\end{pmatrix}
& \quad \text{(For HG)} 
\end{cases}
\end{equation}
and the Gravito-Amp\`{e}re-Maxwell law of MG and HG: 
\begin{equation}\label{eq:83}
\mathbf{\nabla}\times\mathbf{b}\,=\,\begin{cases}
\,-\,\frac{4\pi G}{c^2}\mathbf{j}\,+\,\frac{1}{c^2}\frac{\partial \mathbf{g}}{\partial t} & \quad \text{(For MG)} \\ \\
 \,+\,\frac{4\pi G}{c^2}\mathbf{j}\,-\,\frac{1}{c^2}\frac{\partial \mathbf{g}}{\partial t} & \quad \text{(For HG)} 
 \end{cases}        
\end{equation}
where $\mathbf{b}$ is named as gravitomagnetic field, which is generated by gravitational charge (or mass) current and time-varying gravitational or gravitoelectric field $\mathbf{g}$.\\ 
For reference, we note the field strength tensor with two contravariant indices:
\begin{equation}\label{eq:84}
f^{\alpha \beta} =\eta^{\alpha \gamma}f_{\gamma \delta}\eta^{\delta \beta} = \begin{cases}
\begin{pmatrix}
  0             &   -\frac{g_x}{c}  &   -\frac{g_y}{c}   &  -\frac{g_z}{c} \\
\frac{g_x}{c}   &    0              &   -b_z             &   b_y  \\ 
\frac{g_y}{c}   &    b_z            &    0               &  -b_x  \\ 
\frac{g_z}{c}   &   -b_y            &    b_x             &   0
\end{pmatrix}
& \text{(For MG)} \\ \\ 
\begin{pmatrix}
0    &   -\frac{g_x}{c}  &   -\frac{g_y}{c}   &  -\frac{g_z}{c} \\
\frac{g_x}{c} &    0    &   b_z   &  -b_y  \\ 
\frac{g_y}{c}   &   -b_z  &    0    &  b_x  \\ 
\frac{g_z}{c}   &    b_y  &   -b_x  &   0
\end{pmatrix}
& \text{(For HG)}
\end{cases} 
\end{equation}
From Eq. \eqref{eq:79} and the anti-symmetry property of $f^{\alpha \beta}$, it follows that $j^\alpha$ is divergence-less:
\begin{equation}\label{eq:85}
\partial_\alpha j^\alpha\,=\,0\,=\,\,\frac{1}{c}\frac{\partial (\rho_0c)}{\partial t} + \mathbf{\nabla}\cdot\mathbf{j} = \mathbf{\nabla}\cdot\mathbf{j} \,+\,\frac{\partial \rho_0}{\partial t}.
\end{equation}
This is the {\it continuity equation} expressing the local conservation of proper mass or (proper energy). \\
Equation \eqref{eq:79} gives us two in-homogeneous equations of MG and HG. The very definition of 
$f_{\alpha \beta}$ in Eq. \eqref{eq:67}, automatically guarantees us the Bianchi identity:
\begin{equation}\label{eq:86}
\partial_\alpha f_{\beta \gamma} +\partial_\beta f_{\gamma \delta} + \partial_\gamma f_{\alpha \beta} = 0,
\end{equation}
(where $\alpha , \beta, \gamma $ are any three of the integers $0, 1, 2, 3$), from which two homogeneous equations emerge naturally: 
\begin{equation}\label{eq:87}
\mathbf{\nabla}\cdot \mathbf{b}\,=\,0  \quad \text{(For both MG and HG)}\\
\end{equation}
\begin{equation}\label{eq:88}
\mathbf{\nabla}\times \mathbf{g}= \begin{cases} 
\,-\,\frac{\partial \mathbf{b}}{\partial t} & \quad \text{(For MG)} \\ \\
\,+\,\frac{\partial \mathbf{b}}{\partial t} & \quad \text{(For HG)}
\end{cases} 
\end{equation}
The Bianchi identity \eqref{eq:86} may concisely be expressed by the zero divergence of a dual field-strength tensor $\mathscr{F}^{\alpha \beta}$, viz., 
\begin{equation}\label{eq:89}
\partial _\alpha \mathscr{F}^{\alpha \beta}\,=\,0,
\end{equation} 
where $\mathscr{F}^{\alpha \beta}$ is defined by 
\begin{equation}\label{eq:90}
\mathscr{F}^{\alpha \beta}\,=\,\frac{1}{2}\epsilon^{\alpha \beta \gamma \delta}f_{\gamma \delta}\,=\,\underbrace{\begin{pmatrix}
0     &   -b_x    &    -b_y    &   -b_z \\
b_x   &    0      &    g_z/c   &   -g_y/c  \\ 
b_y   &   -g_z/c  &    0       &    g_x/c  \\ 
b_z   &    g_y/c  &   -g_x/c   &     0
\end{pmatrix}}_{\text{For MG}}
\end{equation}
and the totally anti-symmetric fourth rank tensor $\epsilon^{\alpha \beta \gamma \delta}$ (known as Levi-Civita Tensor) is defined by 
\begin{equation}\label{eq:91}
\epsilon^{\alpha \beta \gamma \delta} =
\begin{cases}
{+1}   & \quad \text{for}\, \alpha = 0, \beta = 1, \gamma = 2, \delta = 3, and \\
{}     & \quad \text{any even permutation}          \\
{-1}   & \quad \text{for any odd permutation}  \\
{}{0}    & \quad \text{if any two indices are equal}. 
\end{cases}
\end{equation} 
The dual field-strength tensor $\mathscr{F}^{\alpha \beta}$ for HG can be obtained from Eq. \eqref{eq:90} by substitution $\mathbf{b} \rightarrow -\mathbf{b}$, with $\mathbf{g}$ remaining the same.\\
In terms of this 4-potentials, 
\begin{equation}\label{eq:92}
{\underline{A}}^\alpha  = (\phi_g/c,\,\mathbf{A}_g),
\end{equation}
the in-homogeneous equations \eqref{eq:79} of MG and HG read:
\begin{equation}\label{eq:93}
\partial_\beta \partial ^\beta A_g^\alpha\,-\,\partial^\alpha(\partial_\beta A_g^\beta)\,=\,-\,\frac{4\pi G}{c^2}j^\alpha.
\end{equation}
\noindent
Under Gravito-Lorenz condition,
\begin{equation}\label{eq:94}
\partial_\beta A_g^\beta =\,0,
\end{equation}
the in-homogeneous Eqs. \eqref{eq:93} simplify to the equations:
\begin{equation}\label{eq:95}
\partial_\beta \partial ^\beta A_g^\alpha\,=\,\Box A_g^\alpha\,=\,-\,\frac{4\pi G}{c^2}j^\alpha \quad\text{(For MG \& HG),}
\end{equation}
where 
\begin{equation}\label{eq:96}
\Box\,=\,\partial_\alpha \partial ^\alpha \,=\,\frac{1}{c^2}\frac{\partial^2}{\partial t^2}\,-\,\mathbf{\nabla}^2
\end{equation}
is the {\it D'Alembertian} operator. In Maxwell's theory of electrodynamics, the equation corresponding to Eq. \eqref{eq:95} is 
\begin{equation}\label{eq:97}
\Box A^\alpha\,=\,\mu_0j_e^\alpha\, \qquad \text{(in SI units)}
\end{equation}
where $\mu_0$ is the permeability of vacuum and the electromagnetic 4-vector potential $A^\alpha$ and the electric 4-current vector $j_e ^\alpha$ are respectively represented by
\begin{equation}\label{eq:98}
A^\alpha = (\phi_e/c,\,\mathbf{A}),\qquad j_e ^\alpha = (c\rho_e ,\, \mathbf{j}_e)
\end{equation}
with the symbols having their usual meanings. The crucial sign difference between the equations \eqref{eq:95} and \eqref{eq:97} will explain why two like masses attract each other under static conditions, while two like charges repel each other under static conditions as we shall see. Since the fundamental field equations are the same for MG and HG, they represent the same physical thing and any sign difference in some particular terms arise due to particular definitions which will not change the nature of physical interactions. Hence, in what follows, what we call MG is to be understood as HMG. \\ 

\section{The Graviton}
In Quantum Gravitodynamics (QGD), $A_g^\alpha$ becomes the wave function of the graviton. Free graviton satisfies Eq. \eqref{eq:95} with $j^\alpha = 0$, 
\begin{equation}\label{eq:99}
\Box A_g^\alpha\,=\,0.
\end{equation}  
If we consider the vacuum plane-wave solutions of Eq. \eqref{eq:99} with four momentum $p = (E/c,\,\mathbf{p})$, then 
\begin{equation}\label{eq:100}
A_g^\alpha(x)\,=\,Ne^{-(i/\hbar)p.x}\epsilon^\alpha(p),
\end{equation}
where $N$ is a normalization factor and $\epsilon^\alpha(p)$ is the polarization vector, which characterizes the spin of the graviton. Substitution of Eq. \eqref{eq:100} into Eq. \eqref{eq:99}, yields a constraint of $p^\alpha$: 
\begin{equation}\label{eq:101}
p^\alpha p_\alpha\,=\,0, \quad {\mbox{or}} \quad E\,=\,|\mathbf{p}|c
\end{equation} 
which is as required for a mass-less particle. \\
Now we notice that $\epsilon^\alpha$ has 4-components, but they are not all independent. The Gravito-Lorenz condition Eq. \eqref{eq:94} demands that 
\begin{equation}\label{eq:102}
p^\alpha \epsilon_\alpha \,=\,0.
\end{equation} 
In the Newton gauge, $\mathbf{\nabla}\cdot\mathbf{A}_g = 0$ (the analogue of Coulomb gauge), we get 
\begin{equation}\label{eq:103}
\epsilon^0 = 0, \quad \mathbf{\epsilon}\cdot \mathbf{p} = 0
\end{equation}
which means that the polarization three vector ($\mathbf{\epsilon}$) is perpendicular to the direction of propagation. So, we say that a free graviton is {\it transversely polarized}. Since there are two linearly independent three-vectors perpendicular to $\mathbf{p}$; for instance, if $\mathbf{p}$ points in the $z$ direction, we might choose 
\begin{equation}\label{eq:104}
\mathbf{\epsilon}^{(1)} = (-1, 0, 0), \quad \mathbf{\epsilon}^{(2)} = (0, -1, 0). 
\end{equation} 
Instead of four independent solutions for a given momentum, we are left with only two. A massive particle of spin $s$ admits $2s +1$ different spin orientations, but a mass-less particle has only two, regardless of its spin (except for $s = 0$, which has only one). Along its direction of motion, it can only have $m_s = +s$\, or $m_s = - s$; its helicity, in other words, can only be $+1$ or $-1$. \\
Thus for a graviton we write
\begin{equation}\label{eq:105}
A_{g\alpha}(x)\,=\,Ne^{-(i/\hbar)p.x}\epsilon_\alpha^{(s)}
\end{equation}
where $s = 1, 2$ for two spin states (polarizations). The polarization vectors $\epsilon_\alpha ^{(s)}$ satisfy the momentum space Gravito-Lorenz condition \eqref{eq:102}. They are orthogonal in the sense that 
\begin{equation}\label{eq:106}
\epsilon_\alpha^{(1)*}\epsilon^{(2)\alpha} = 0
\end{equation}
and normalized 
\begin{equation}\label{eq:107}
\epsilon^{\alpha *}\epsilon_\alpha \,=\,-\,1.
\end{equation}
In the Newton gauge, $\mathbf{\nabla}\cdot\mathbf{A}_g = 0$, the polarization three-vectors obey the {\it completeness} relation 
\begin{equation}\label{eq:108}
\sum _{s = 1,2}\epsilon_i^{(s)}\epsilon_j^{(s)*}\,=\,\delta_{ij}\,-\,\hat{p_i}\hat{p_j}.
\end{equation}
Regarding the idea of spin-2 graviton, Wald (\cite{31}, pp.76) noted that the linearized Einstein's equations in vacuum are precisely the equations written down by Fierz and Pauli \cite{58}, in 1939, to describe a massless spin-2 field propagating in flat space-time. {\it Thus, in the linear approximation, general relativity reduces to the theory of a massless spin-2 field which undergoes a non-linear self- interaction. It should be noted, however, that the notion of the mass and spin of a field require the presence of a flat back ground metric $\eta_{ab}$ which one has in the linear approximation but not in the full theory, so the statement that, in general relativity, gravity is treated as a mass-less spin-2 field is not one that can be given precise meaning outside the context of the linear approximation} \cite{31}. Even in the context linear approximations, the original idea of spin-2 graviton gets obscured due to the several faces of non-isomorphic Gravito-Maxwell equations seen in the literature from which a unique and unambiguous prediction on the spin of graviton is difficult to get as shown in Ref. \cite{43}.
\section{Attraction Between Like Masses}
Let us find the static interaction between two point (positive) masses at rest, following a classical approach \cite{59} within the framework of Maxwellian Gravity as follows.
For a particle having gravitational charge $m_g = m_0$ at rest at the origin, the 4-current densities can be shown to be \cite{59}:
\begin{equation}\label{eq:109}
j^0 = m_0c\delta^3(\mathbf{x}),\qquad \mathbf{j} = \mathbf{0}.
\end{equation}
In Eq. \eqref{eq:95}, we can therefore set 
\begin{equation}\label{eq:110}
A_g^0 = \phi_g/c, \qquad  \mathbf{A}_g = \mathbf{0},
\end{equation}
where 
\begin{equation}\label{eq:111}
\mathbf{\nabla}^2\phi_g = 4\pi G m_0 \delta^3(\mathbf{x}).
\end{equation}
This is nothing but the Poisson's equation for gravitational potential of a point mass at rest at origin. Using Green's Function, the potential at a distance $r$ for a central point particle having gravitational mass $m_0$ (i.e., the fundamental solution) is
\begin{equation}\label{eq:112}
\phi_g(r) = -\frac{Gm_0}{r},
\end{equation}
which is equivalent to Newton's law of universal gravitation. The interaction between two point particles having gravitational charges $m_0$ and $m_0^\prime$ separated by a distance $r$ is 
\begin{equation}\label{eq:113}
U_{12} = m_0^\prime \phi_g = - \frac{G m_0^\prime m_0}{r},
\end{equation}
which is {\it negative} for like gravitational charges and {\it positive} for un-like gravitational charges, if they exist. With $m_0$ at rest at the origin (designated as mass 1), the force on another stationary gravitational charge $m_0^\prime$ (designated as mass 2) at a distance $r$ from origin is 
\begin{equation}\label{eq:114}
\mathbf{F_{21}} = - m_0^\prime \mathbf{\nabla}\phi_g(r)= -\frac{Gm_0 m_0^\prime}{r^2}\mathbf{\hat{r}} = - \mathbf{F_{12}}.
\end{equation}
This force is attractive, if $m_0$ and $m_0^\prime$ are of same sign and repulsive if they are of opposite sign - the reverse case of electrical interaction between two static electric charges. \\
In stead of the above classical approach, one may follow Feynman's \cite{2} detailed quantum field theoretical approach using our Eq. \eqref{eq:95} to arrive at our above conclusion. This is possible because of a difference in sign as seen in the Eqs. \eqref{eq:95} and \eqref{eq:97} here. Zee's \cite{3} path-integral approach may also be used to arrive at the same conclusion. The fact that a vector gravitational theory can produce attractive interaction was transparently clear to Sciama \cite{33}.
\subsection{Lagrangian For Quantum Gravitodynamics}
According to the present study, the final expression for the Lagrangian density for quantum gravitodynamics (QGD) of neutral massive Dirac fields interacting with fields of Maxwellian Gravity (spin-1 gravitons) in flat space-time turns out (in SI units) as 
\begin{equation}\label{eq:115}
\mathcal{L}_{QGD} = [i\hbar c\overline{\psi}\gamma^\mu\partial_\mu \psi - m_0c^2\overline{\psi}\psi]
-\frac{c^2}{16\pi G}f_{\mu \nu}f^{\mu \nu}+ j^\mu A_{g\mu}, 
\end{equation}
where $j^\mu = m_0c(\overline{\psi}\gamma^\mu\psi)$ and $A_{g\mu}$ are the solutions of the  Eq. \eqref{eq:95}. Note that the sign of the free-field terms $f_{\mu \nu}f^{\mu \nu}$ determine the sign of the fee Hamiltonians (or, better, energy densities) being positive and definite in Eq. \eqref{eq:115}. \\
In this work, the gravitational energy density for free fields is fixed positive by choice to address the objection of MTW (Sec. 7.2)\cite{5} without any inconsistency with the field equations of MG. Our quantum field theoretical derivation of MG (assuming the positive energy carried by freely propagating fields) corroborates all the suggested or derived linear vector gravitational equations in flat space-time reviewed in sect. 2. This means that the field equations of MG have rooms for both positive and negative energy solutions. This is because, the Lagrangian density for a particular system is not unique; one can always multiply $\mathcal{L}$ by a constant, or add a constant - or for that matter the divergence of an arbitrary function ($\partial_\mu M^\mu$,\, where $M^\mu$ is any function of $\phi_i$ and $\partial_\mu \phi_i$); such terms cancel out when we apply the Euler-Lagrange equations, so they do not affect the field equations \cite{60}. For instance, we can multiply equation \eqref{eq:49}  by $-1$ to obtain another Lagrangian density $\mathcal{L}_{MG}^\prime = - \mathcal{L}_{MG}$, which  would imply negative energy for the fields - probably these fields are static (non-propagating) fields of gravitostatics/gravito-magnetostatics. However, the issue of this positive vs negative energy solutions and their physical interpretations/implications in the context of gravitation is far from clear and is being investigated by the authors.  \\

\section*{Conclusions}
Following two independent approaches involving (a) the Lorentz-invariance of physical laws under special relativity and (b) the principle of local gauge invariance of quantum field theory applied to a massive electrically neutral Dirac spin $1/2$ particle, we rediscovered spin-1 Heaviside-Maxwellian Gravity, in which like masses are shown to attract each other under static conditions, contrary to the standard view of field theorists. We also suggested a Lagrangian density in which the freely propagating gravito-electromagnetic fields carry positive energy. Here, we for the first time corrected Heaviside's speculative gravito-Lorentz force law. The theory looks interesting and important, particularly in respect of its quantization and unification with other fundamental forces of nature. It may shed some new light on our understanding of the nature of physical interactions and their interplay at the quantum level. We further we note that HMG is valid only in inertial frames, the existence of which is not denied by the General Relativity, which is silent on the nature of gravity in inertial frames. Appropriate modifications are necessary to extend HMG to the case of non-inertial frames.
%
%



\begin{thebibliography}{}
\bibitem{12} An unedited copy of the original Heaviside's article, except that some formulas and all vector equations have been converted to modern notation, is reproduced in \cite{13} below, p. 189-202.
\bibitem{50} D. H. Menzel, W. W. Salisbury, Nature {\bf 257} 162 (1975). This is a response to Lorrain \cite{48}. Present authors are not satisfied with this response because Menzel and Salisbury did not prove $m_g = m = m_0/\sqrt{1- v^2/c^2}$, although they said $m = m_0/\sqrt{1- v^2/c^2}$ has the experimental validity.
\end{thebibliography}


\end{document}